\documentclass[twocolumn,showpacs,preprintnumbers,amsmath,amssymb]{revtex4}
\usepackage{tabularx,graphicx}\begin{document}
\newcommand{\beq}{\begin{equation}}
\newcommand{\eeq}{\end{equation}}
\newcommand{\beqn}{\begin{eqnarray}}
\newcommand{\eeqn}{\end{eqnarray}}
\newcommand{\bmath}{\begin{subequations}}
\newcommand{\emath}{\end{subequations}}
\title{Prediction of unexpected behavior of the mean inner potential of superconductors}
\author{J. E. Hirsch }
\address{Department of Physics, University of California, San Diego\\
La Jolla, CA 92093-0319}
 
\date{\today} 
\begin{abstract}

Off-axis electron holography can measure the mean inner electric potential of materials. The theory of hole superconductivity predicts that 
when a material is cooled into the superconducting state it expels   electrons from its interior to the surface, giving rise to 
a mean inner potential that increases with sample thickness. Instead, in a normal metal and in a conventional BCS superconductor the mean inner potential is expected to be independent
of sample thickness and temperature. Thus, this experiment can provide a definitive test of the validity of the theory of hole superconductivity.
\end{abstract}
\maketitle 

Off-axis electron holography   measures the interference of a reference electron wavefront propagating in vacuum with 
one propagating through a material\cite{hg1,hg2,hg0}. 
Conceptually it is simply Young's  double slit interference experiment with electron waves where one of the 'slits' contains the material to be studied. The wave passing through the material
undergoes a phase shift that depends on the electrostatic and magnetostatic fields in the material. Thus, the interference pattern provides direct information on the electric and magnetic fields
and potentials in the sample. The lateral spatial resolution of the technique is a few nm, and sample thicknesses up to 500nm can be studied with electron beam energies of order 
hundreds of $keV$'s, yielding electric potential resolution better than
$0.1V$\cite{hg1}. These characteristics make it ideal for the problem of interest here.

\begin{figure}
\resizebox{8.5cm}{!}{\includegraphics[width=7cm]{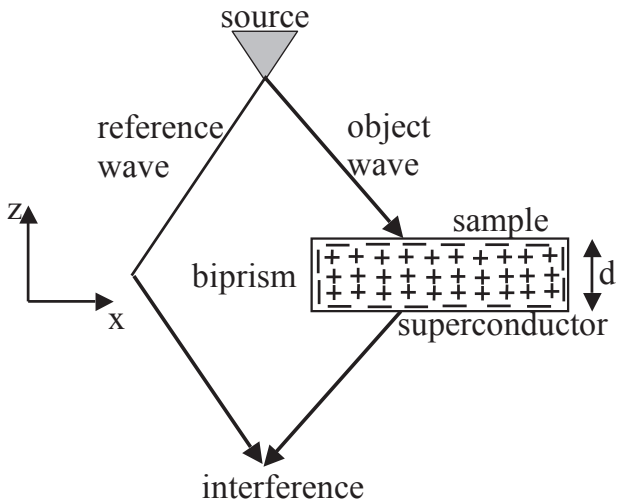}}
  \caption{ Schematic depiction of off-axis electron holography experiment. The object wave traversing the superconducting sample will advance its phase due to the presence of an
  additional positive potential in the superconductor. }
\end{figure} 

The theory of hole superconductivity\cite{hole1} predicts that electrons are expelled from the interior of the sample to a surface layer of thickness given by the London penetration depth when a material goes
superconducting, thus giving rise to an electrostatic field in the interior. This charge expulsion is a key component of the theory and intimately related to many other aspects of the theory, in particular 
it is at the heart of the explanation of the Meissner effect within this theory\cite{meissner}, the prediction that macroscopic spin currents exist in the ground state of superconductors\cite{sm}, and the
prediction that superconductivity is kinetic energy driven\cite{kinetic}. 
Instead, within conventional BCS theory no such charge expulsion nor spin currents exist in superconductors, no explanation of the Meissner effect exists\cite{meissner},  and superconductivity is potential energy driven. 
In this paper we  point out  that the technique of off-axis electron holography should be able to definitely confirm or rule out the charge expulsion predicted by the theory of hole
superconductivity, thus strongly supporting or ruling out the theory.

Figure 1 shows the conceptually very simple experimental setup. The superconducting slab of thickness $d$ is predicted to have excess negative charge in the regions within 
a distance $\lambda_L$, the London penetration depth, of the surfaces and excess positive charge in the deep interior. The maximum electric field in the interior, denoted by $E_m$, is predicted to be given by the lower
critical field $H_{c1}$\cite{electrospin} (e.g. $H_{c1}=200G$ corresponds to $E_m=60,000V/cm$). The electric field goes to zero as one approaches the boundaries of the sample.

The phase change of the electron wave passing through a sample slab relative to the wave propagating in vacuum is given by\cite{hg1}
\beq
\phi(x)=C_E\int V(x,z)dz
\eeq
with
\beq
C_E=\frac{2\pi |e|}{\lambda E}\frac{E+m_e c^2}{E+2m_e c^2}
\eeq
with $E$ the electron kinetic energy in vacuum, $\lambda=hc/\sqrt{E^2+2Em_ec^2}$ its wavelength and $m_e$ and $e$ its mass and charge respectively. $z$ is the propagation direction perpendicular to
the slab and $x$ is the horizontal direction. In a normal metal the electrostatic potential $V(x,z)$ is expected to be approximately constant and Eq. (1) is simply, assuming uniform
thickness $d$
\beq
\phi =C_E \bar{V}_0d
\eeq
with $\bar{V}_0$ termed the ``mean inner potential''\cite{mip1,mip2,mip3} which is a characteristic of the material, typically between $5$ and $30$ Volts
 (positive). Thus, for the superconductor in the normal state the phase shift Eq. (3) is directly
proportional to the thickness of the sample $d$. The expected  linear dependence of the phase shift on  sample thickness in electron holography experiments with non-superconductors has been  verified experimentally
for a variety of materials\cite{mip2}.

 \begin{figure}
\resizebox{8.5cm}{!}{\includegraphics[width=7cm]{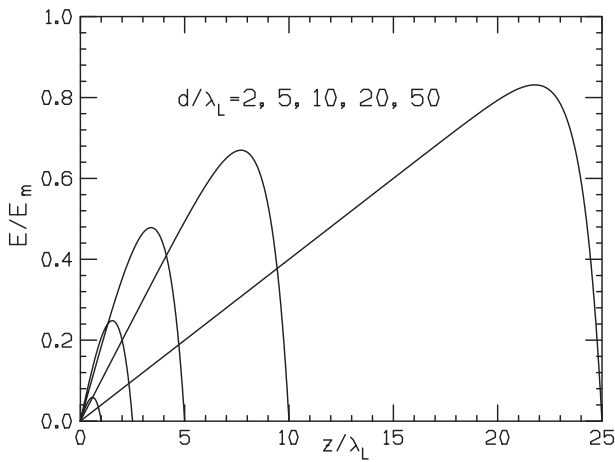}}
  \caption{Electric field resulting from charge expulsion
  from the center of the sample ($z=0$) to the upper edge ($z=d/2$)  for samples of various thicknesses $d$, far from the lateral edges of the sample. The electric
  field points in the $+z$ direction. }
\end{figure}

According to the theory of hole superconductivity the charge density in the interior of superconductors satisfies the differential equation\cite{chargeexp}
\beq
\rho(\vec{r})=\rho_0+\lambda_L^2\nabla^2 \rho(\vec{r})
\eeq
with $\rho_0$ a positive constant denoting a positive charge density deep in the interior of the superconductor. Eq. (4) and the condition of overall charge neutrality predict that there is excess
negative charge within a London penetration depth of the surfaces of the sample. The resulting electrostatic field $\vec{E}(\vec{r})$ satisfies the equation
\beq
\vec{E}(\vec{r})=\vec{E}_0(\vec{r})+\lambda_L^2\nabla^2\vec{E}(\vec{r})
\eeq
with $\vec{E}_0(\vec{r})$ the electrostatic field resulting from a uniform charge densith $\rho_0$ throughout the superconductor. Deep in the interior $\vec{E}(\vec{r})=\vec{E}_0(\vec{r})$.
The value of $\rho_0$ is determined by the condition that the electric field approaches a maximum value $E_m$ near the surface 
of a sample of dimensions much larger than $\lambda_L$, with\cite{electrospin}
\beq
E_m=-\frac{\hbar c}{4e\lambda_L^2}
\eeq
i.e. essentially the lower magnetic critical field $H_{c1}$\cite{tinkham}.

In an infinite slab of thickness $d$ with normal in the $z$ direction and centered at $z=0$ the electric field points in the $\pm z$ direction and is given by\cite{electrospin}
\beq
E(z)=\frac{2 E_m z}{d}  (1-\frac{d}{2z} \frac{sinh(\frac{z}{\lambda_L})}{sinh(\frac{d}{2\lambda_L})}) .
\eeq
Figure 2 shows the electric field as function of $z$ for samples of increasing thickness for fixed $\lambda_L$. Note that even for $d/\lambda_L=50$ the maximum electric
field is only about $0.8$ of its limiting value $E_m$.

The electric potential in the slab arising from charge expulsion is given by
\beqn
V_{ce}(z)&=&\frac{E_m d}{4}(1-\frac{4z^2}{d^2})  + \nonumber \\
& &\frac{E_m \lambda_L}{sinh(\frac{d}{2\lambda_L})} [cosh(\frac{z}{\lambda_L})-cosh(\frac{d}{2\lambda_L})]   
\eeqn     
and its contribution the mean inner potential, defined by 
 \beq
 \bar{V}_{ce}=\frac{2}{d}\int_0^{d/2} V_{ce}(z)dz
 \eeq
 is given by
 \beq
 \bar{V}_{ce}=\frac{E_md}{6} + 2E_m\frac{\lambda_L^2}{d} -E_m\lambda_L    \frac{cosh(\frac{d}{2\lambda_L})}{sinh(\frac{d}{2\lambda_L})}
 \eeq
 This potential should be added to the ordinary mean inner potential $\bar{V}_0$ arising from the local electronic charge distribution in the unit cell\cite{mip1}, which is independent of
 sample thickness.
 For a slab of thickness $d$ the phase shift is then
 \bmath
 \beq
  \phi=C_E(\bar{V}_0 d +  \bar{V}_{ce}d)
 \eeq
 and in particular if $d$ is much larger than the London penetration depth the first term in Eq. (10) dominates and the phase shift is
 \beq
 \phi=C_E(\bar{V}_0 d +  \frac{E_m}{6}d^2)
 \eeq
 \emath
 that is, it has a linear and a quadratic contribution in  the slab thickness $d$, in contrast to the purely linear behavior  expected in a normal metal.

 \begin{figure}
\resizebox{8.5cm}{!}{\includegraphics[width=7cm]{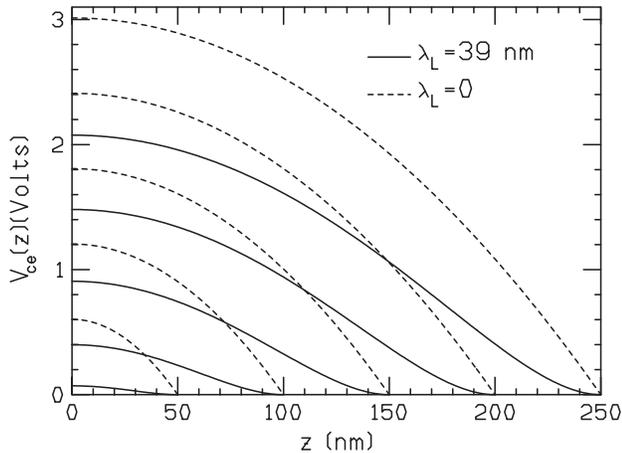}}
  \caption{Electric  potential resulting from charge expulsion for $z$ ranging from the center of the sample ($z=0$) to the upper edge ($z=d/2$)  for samples of thicknesses 
  $d=100, 200, 300, 400, 500 nm$, far from the lateral edges of the sample, for $\lambda_L=39nm$ (solid lines) and $\lambda_L=0$ (dashed lines). 
  The electric potential goes to zero at the upper edge of the sample ($z=d/2$). The magnitude of the electric potential corresponds to the case of 
  $Pb$ (see text).}
\end{figure}

Let us consider for definiteness a $Pb$ sample. The London penetration depth at low temperatures is $\lambda_L=39 nm$.
There is approximately one excess electron every $10^6$ atoms near the surface, and  
$E_m=0.0241 V/nm$\cite{electrospin}. Figure 3 shows the electric potential arising from charge expulsion
as function of $z$ from the center to the top (or bottom) of the slab for samples of varying thicknesses. The potential goes to
zero at the surface of the sample and is maximum at the center. The difference between the solid and corresponding dashed lines in Fig. 3 illustrates the effect of
the finite $\lambda_L$.

Figure 4 shows the contribution to the mean inner potential arising from charge expulsion 
Eq. (10) as a function of sample thickness assuming the value of $E_m$ for $Pb$ and various values for the London penetration
depth. In the limit of small $\lambda_L$ the dependence on $d$ is linear as given by the first term in Eq. (10). As $\lambda_L$ increases, the behavior becomes nonlinear and the
magnitude decreases.

Such inner potentials should give rise to easily detectable phase shifts in an electron holography experiment. 
For example, in an experiment with $300keV$ electrons the constant $C_E$ is $0.0065 (nm)^{-1} V^{-1}$ and the contribution to the mean inner potential arising
from charge expulsion for a $500nm$ thick $Pb$ sample is predicted to be $1.22V$ (Fig. 4), giving rise to an additional phase shift of $3.5$ radians.
The non-linear dependence of the phase shift
$\phi$ on the sample thickness should provide direct evidence for the physics discussed here, and should be fittable with the formulas given here with a value
of $\lambda_L$ that agrees with the value of the London penetration depth obtained independently from magnetic measurements.

 \begin{figure}
\resizebox{8.5cm}{!}{\includegraphics[width=7cm]{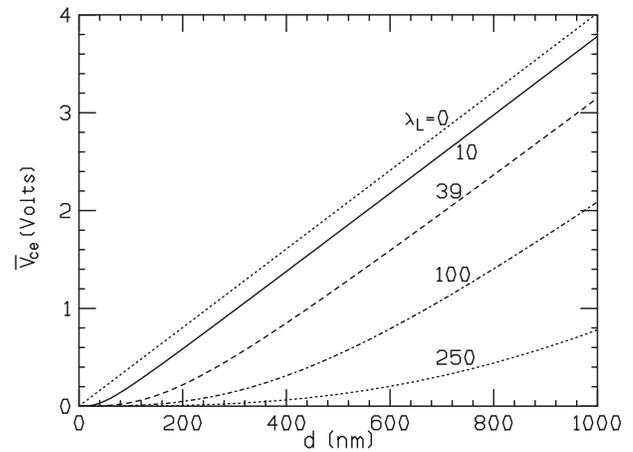}}
  \caption{ Mean inner potential resulting from charge expulsion as a function of sample thickness $d$ for various values of the London penetration depth
  (numbers next to the lines, in $nm$). The value of the maximum electric field corresponds to the case of $Pb$, and $\lambda_L=39nm$ corresponds to $Pb$. The total mean inner potential
  has in addition a thickness-independent contribution which is the same as in the normal state. }
\end{figure} 

The expression Eq. (10) for the inner potential arising from charge expulsion applies for electron beam  paths far away from the lateral edges of the sample.
Approaching a lateral edge, the internal electric field direction changes and points towards the lateral surface when the distance to the lateral surface becomes smaller than
$d/2$ (roughly speaking the internal electric field arising from charge expulsion points towards the closest surface\cite{chargeexp}). From that point on the contribution to the inner
potential from this physics starts to decrease rapidly and even more so as the distance to the lateral edge becomes smaller than the London penetration depth.
Thus, a mapping of the phase shift as a function of $x$ and $d$ should yield detailed information to check the theoretical predictions. Other sample geometries may
also be useful, for example the wedge geometry used in Ref. \cite{mip2}. The electric potential and the mean internal potential arising from charge expulsion
for samples of arbitrary shape can be calculated by numerical solution of the differential equations describing the electrodynamics of the superconducting
state within this theory\cite{chargeexp}.

Next we need to examine at what temperatures will these effects be observable. We can think of the superconductor at finite temperature as a mixture of superfluid and
normal fluid. According to the theory of hole superconductivity superfluid is expelled to the surface, however this effect will be countered to some extent by a backflow of
normal fluid to attempt to preserve charge neutrality. We can estimate the effect of temperature using a simple two-fluid description. In a BCS superconductor the normal fluid density at finite
temperature is given by\cite{tinkham}
\beq
n_n(T)=2n_s\int_\Delta^\infty dE (-\frac{\partial f}{\partial E}) \frac{E}{\sqrt{E^2-\Delta^2}}
\eeq
where $n_s$ is the superfluid density at zero temperature, $\Delta$ is the superconducting energy gap and $f(E)$ the Fermi function. This expression applies also approximately to the
model of hole superconductivity\cite{london}. At low temperatures we can approximate Eq. (12) by
\beq
n_n(T)= (2\pi \beta \Delta)^{1/2}e^{-\beta \Delta}
\eeq
with $\beta=1/k_BT$. Assuming the normal particles carry a full electron charge, the normal fluid will not be sufficient to screen the positive charge $\rho_0$ in the interior resulting from the
superfluid expulsion when the condition
\beq
|e|n_n(T)<\rho_0
\eeq
is satisfied. According to the theory\cite{electrospin}
\beq
\rho_0=2\rho_-\frac{\lambda_L}{d}
\eeq
with $\rho_-$ the density of negative charge near the surface, given by\cite{electrospin}
\beq
\rho_-=\frac{r_q}{2\lambda_L}en_s
\eeq
with $r_q=\hbar/(2m_ec)=0.00193\AA$. Therefore the condition Eq. (14) is
\beq
n_n(T)<\frac{r_q}{d}n_s
\eeq
or, using Eq. (13)
\beq
(\beta \Delta)^{1/2} e^{-\beta\Delta}<\frac{r_q}{\sqrt{2\pi}d} .
\eeq
Using $2\Delta/k_B T_c=3.53$ with $T_c$ the critical temperature, and $d=500nm$ yields 
\beq
T<0.10T_c
\eeq
as a necessary condition for the internal electric field not to be screened by the normal quasiparticles. For $Pb$ with $T_c=7.193^oK$ this would require cooling the sample to about $0.7^oK$. For smaller values of $d$, Eq. (18) suggests that the required temperature is higher, however there are additional corrections and
one finds that the required temperature is within $5\%$ of Eq. (19) in the entire range $50nm<d<500nm$. 

In reality we believe that the condition Eq. (19) is much too stringent, because it was obtained assuming that the normal quasiparticles carry a full electron charge. In a BCS superconductor
in fact quasiparticles are exactly charge neutral on average. Within the theory of hole superconductivity quasiparticles carry a positive charge on average, but it is much smaller than
one electron charge\cite{thermo}. Thus we argue that the condition Eq. (19) can certainly be expected to be sufficient for the effects predicted here to be seen,
and in fact the  effects may show up already at substantially higher   temperature.

As discussed earlier, a dependence of the mean inner potential on the thickness of the sample and the position $x$ of the beam with respect to the edge of the sample should not be seen
in the normal state. Such effects should be seen in superconductors at sufficiently low temperatures according to our theory. For
 given  sample thickness and position $x$ a substantial increase in phase shift will be seen when the temperature is
lowered sufficiently below $T_c$ and the internal electric field resulting from charge expulsion becomes unscreened. In addition, the 
increase in  phase shift can be reversed by application of a magnetic field in the $z$ direction larger than the critical
field, that would render the system normal and undo the charge expulsion. None of these effects should be seen in a normal metal nor in a conventional
BCS superconductor: in those systems the phase shift will be independent of sample thickness and position $x$, independent of temperature, and would
not change under application of a magnetic field in the $z$ direction (it would with a magnetic field in the in-plane direction). 

In summary: the theory of hole superconductivity predicts that superconductors expel electrons from the interior to the surface, and that 
as a consequence a macroscopic  electric field and resulting electric potential exist in the interior
of superconductors at sufficiently low temperatures. The conventional theory of superconductivity does not predict this behavior. Off-axis electron holography can test this prediction and render 
unambiguous experimental evidence for or against  it. The experiment can be performed with any superconductor, since the theory 
 is predicted to apply to all
superconductors.  Clear experimental evidence for any superconductor that electrons are $not$ expelled from the interior to the surface in the superconducting state would
falsify the theory of hole superconductivity. On the other hand, the opposite experimental result would falsify BCS theory only for that particular material, since BCS theory is not expected to applyÊ
to      all superconducting materials\cite{cohen}.

 \end{document}